\documentstyle{article}

\begin{document}
\large

\begin{center}
{\LARGE Quantum key distribution based on Greenberger-Horne-Zeilinger state}\\[1.0cm]
Guihua Zeng\\[0.1cm]
National Key Lab. on ISDN, XiDian University, Xi'an 710071, P.R.China\\[1.5cm]

{\Large Abstract}\\[0.2cm]
\end{center}

{\small  An unsymmetrical quantum key distribution scheme is proposed, its security is 
guaranteed by the correlation of the Greenberger-Horne-Zeilinger triplet state.
In the proposed protocol, the distribution of quantum states are unsymmetrical.
This unsymmetrical characteristic makes the transmission qubits (except the loss qubits) 
be useful. Sequentially 
the proposed protocol has excellent efficiency and security which are very useful in the practical application. }

Quantum key distribution 
is defined as a procedure allowing two (multi-) legitimate users of communication channel
to establish two (multi-) exact copies, one copy for each user, of a random and secret
sequence of bits. It employs quantum phenomena such as the Heisenberg uncertainty 
principle and the quantum corrections to protect distributions of cryptographic keys. 
Since the BB84 protocol was presented, lots of quantum key distribution protocols for two communicators have been proposed. Three main protocols are the BB84 protocol [1], B92 
protocol [2] and the EPR protocol [3,4]. 
All these presented quantum key distribution protocols are provably secure against 
eavesdropping attacks [5-9],
in that, as a matter of fundamental principle, the secret data can not be
compromised unknowingly to the legitimate users of the channel. Currently, the quantum
key distribution has entered the experimental phases. 
The first quantum key distribution prototype, working over a distance 
of 32 centimeters in 1989, was implemented by means of laser transmitting in free space [9].
Soon, experimental demonstrations by optical fibber were set up [10]. 
Recently, the transmission distance is extended to more than 30Km in the fiber [11] 
and 205m in the free space [12].

The first quantum key distribution scheme, i.e., the Bennett Brassard
(BB84) scheme, was presented a decade ago. It is implemented by the four states $\{|\uparrow\rangle, 
|\downarrow\rangle, |\nearrow\rangle, |\searrow\rangle \}$, where any of the two states $\{|\uparrow\rangle, |\downarrow\rangle\}$ and any of the two states $\{|\nearrow\rangle, |\searrow\rangle\}$ are non-commuted. Its security relies on the uncertainty principle of 
quantum mechanics. The security guarantee is derived from
the fact that each bit of data is encoded at random on either one of a
conjugate pair of observables of quantum-mechanical object. Because such a
pair of observables is subjected to the Heisenberg uncertainty principle,
measuring one of the observables necessarily randomizes the other. In 1992, Bennett devised another protocol, i.e., the B92 protocol. It is based on the transmission of nonorthogonal
quantum states. This protocol uses any two nonorthogonal states to implement the quantum key
distribution. Its security relies on the non-distinguishability of unknown 
two-nonorthogonal states.

Recently another quantum key distribution scheme, based on Einstein-Podolsky-Rosen 
(EPR) [13] correlations, was suggested by Ekert and modified by Bennett, Brassard, and 
Mermin. In the Ekert's vision, it uses the EPR pair to distribute the quantum 
cryptographic key, and 
uses the violation of the Bell inequalities to provide the secret security. In the Bennett, Bassard and Mermin's revised vision, they use the EPR pair to distribute the quantum cryptographic key, but use the EPR correlation to provide the secret security. We describe
here the modified version. In this scheme the communicator, called Alice, creats pairs of 
spin 1/2 particles in the 
singlet state, and sends the communicator, called Bob, one particle from each pair. When the two particles are 
measured separately the results obtained for them are correlated. Using the correlation
Alice and Bob can check the eavesdropping and obtain the sharing key.

In this letter, we present another method in which the communicators use the 
Greenberger-Horne-Zeilinger (GHZ) state [14] to
distribute the quantum cryptographic key and use the correlations of the GHZ triplet state to 
provide the secret security. In this scheme two particles of the GHZ triplet state are 
distributed to Alice and the third particle of GHZ triplet state is distributed to Bob. 
Obviously, the distribution of the particles and the quamtum states are unsymmetrical in the proposed protocol, which is different from the previous schemes in which Alice and Bob the 
same numbers particles or quantum states. This unsymmetrical characteristic brings some merits
than the previous protocol, for example the efficiency approaches $100\%$ and the security is 
more higher, these characteristics are very useful in the practical application. In the proposed
protocol we let Alice or the quantum source create a sequence of GHZ triplet states, 
and send one particle from each triplet state to Bob. When three particles are 
measured the results obtained for them are correlated. By using the correlation of the
GHZ triplet state Alice and Bob can check the eavesdropping and obtain the sharing key.

Before describing our protocol, we first investigate the correlation of the GHZ triplet state.
In general, A N-particle entanglement states may be written as 
\begin{equation}
|\psi\rangle=\prod^N_{i=1}|u_i\rangle\pm \prod^N_{i=1}|u^c_i\rangle,
\end{equation}
where $u_i$ stand for a binary variable, $u_i\in \{|z+\rangle, |z-\rangle\}$ and $u^c_i=1-u_i$, 
$|z+\rangle$ and $|z-\rangle$ denote the spin eigenstates, or equivalently the horizontal and 
vertical polarization eigenstates, or equivalently any two-level system.
For $N=2$ they reduce to the Bell states and $N=3$ and $N=4$ they represent the GHZ 
states. For a general $N$ we shall calling them cat states. In this paper, we are 
interesting in the case of $N=3$, i.e., the GHZ triplet state.

In Eq.(1) for $N=3$ they reduce to eight GHZ triplet states, we here use the following state

\begin{equation}
|\psi\rangle=\frac{1}{\sqrt{2}}(|z+z+z+\rangle+|z-z-z-\rangle).
\end{equation}
Define the $x$ and $y$ eigenstates

\begin{equation}
|x+\rangle=\frac{1}{\sqrt{2}}(|z+\rangle+|z-\rangle),
\end{equation}

\begin{equation}
|x-\rangle=\frac{1}{\sqrt{2}}(|z+\rangle-|z-\rangle),
\end{equation}

\begin{equation}
|y+\rangle=\frac{1}{\sqrt{2}}(|z+\rangle+i|z-\rangle),
\end{equation}

\begin{equation}
|y-\rangle=\frac{1}{\sqrt{2}}(|z+\rangle-i|z-\rangle),
\end{equation}
the GHZ triplet state can be rewritten as 
\begin{equation}
\begin{array}{rl}
|\psi\rangle=&\frac{1}{2}[(|x+\rangle|x+\rangle+|x-\rangle|x-\rangle)|x+\rangle\\
             & \\ 
             &+(|x+\rangle|x-\rangle+|x-\rangle|x+\rangle)|x-\rangle],
\end{array}
\end{equation}
or 
\begin{equation}
\begin{array}{rl}
|\psi\rangle=&\frac{1}{2}[(|y+\rangle|y-\rangle+|y-\rangle|y+\rangle)|x+\rangle\\
             & \\
             &+(|x+\rangle|x-\rangle+|x-\rangle|x+\rangle)|x-\rangle],
\end{array}
\end{equation}
or 
\begin{equation}
\begin{array}{rl}
|\psi\rangle=&\frac{1}{2}[(|y+\rangle|x-\rangle+|y-\rangle|x-\rangle)|y+\rangle\\
             & \\
             &+(|y+\rangle|x+\rangle+|y-\rangle|x-\rangle)|y-\rangle],
\end{array}
\end{equation}
or 
\begin{equation}
\begin{array}{rl}
|\psi\rangle=&\frac{1}{2}[(|x+\rangle|y-\rangle+|x-\rangle|y+\rangle)|y+\rangle\\
             & \\
             &+(|x+\rangle|y+\rangle+|x-\rangle|y-\rangle)|y-\rangle].
\end{array}
\end{equation}
The above decomposition demonstrates the correlation among three particles. For example, in 
Eq.(7) if one particle is in the state $|x+\rangle$ and the second particle is in the state
$|x+\rangle$, the third particle must be in the state $|x+\rangle$ because of the correlation 
of the GHZ triplet state. By Eqs.(7-10), one may construct a lock-up table to summarize these properties of GHZ states. 

\begin{center}
Table I. The correlation results of the GHZ triplet states\\[0.1cm] 

\begin{tabular}{c|llll}
\hline
\hline
Particle 1  &$|x+\rangle$ &$|x-\rangle$ &$|y+\rangle$ &$|y-\rangle$\\
\hline
Particle 2 &$|x+\rangle$  &$|x+\rangle$  &$|x+\rangle$ &$|x+\rangle$ \\
Particle 3 &$|x+\rangle$  &$|x-\rangle$  &$|y-\rangle$ &$|y+\rangle$ \\
\hline
Particle 2 &$|x-\rangle$  &$|x-\rangle$  &$|x-\rangle$ &$|x-\rangle$ \\
Particle 3 &$|x-\rangle$  &$|x+\rangle$  &$|y+\rangle$ &$|y-\rangle$ \\
\hline
Particle 2 &$|y+\rangle$  &$|y+\rangle$  &$|y+\rangle$  &$|y+\rangle$ \\
Particle 3 &$|y-\rangle$  &$|y+\rangle$  &$|x-\rangle$  &$|x-\rangle$ \\
\hline
Particle 2 &$|y-\rangle$  &$|y-\rangle$  &$|y-\rangle$  &$|y-\rangle$ \\
Particle 3 &$|y+\rangle$  &$|y-\rangle$  &$|x+\rangle$  &$|x-\rangle$ \\
\hline
\hline
\end{tabular}
\end{center}

In any column of the table I we assume the result of particle 1 to be determined, and 
consider the possible results of particles 2 and 3. The table I shows two properties of 
the GHZ triplet state: 
i) By the result of any of the three particles, one can determine whether the other two 
results of the particles are the 
same or not and also that he (she) will gain no knowledge of what their 
results actually are, if he (she) knows what measurements have been made for  
the other two particles (that is $x$ or $y$). 
ii) From table I it is clear that allows two parties jointly, but only jointly, 
to determine which was the measurement outcome of the third party. So if 
the measurement directions of 
the three participators are public, the combined results of any two participators 
can determine what the result of the third party's measurement was. 

As discussed above, the GHZ state has a correlation properties that if only one particle 
has been measured, the states of the other two particles
are still not determined. This property may be used in the quantum key distribution
scheme. Let us now show how to implement the quantum key distribution scheme by the GHZ state. 
Suppose Alice have two particles denoted by $P_1, P_3$, and Bob have 
one particle denoted by $P_2$ from the GHZ triplet state. The protocol goes as follows.

{\it 
\begin{enumerate}

\item Alice makes a random measurement on her one particle (e.g., $P_1$) of two GHZ particles, either in the $x$ or $y$ direction. 

\item Bob makes a random measurement on his GHZ particle $P_2$, either in the $x$ or $y$ 
direction. 

\item Bob sends his measurement bases $x$ or $y$ to Alice, but not the qubit values.

\item Alice measures his particle $P_3$ according to the measurement bases of the particles 
$P_1$ and $P_2$.

\item Check eavesdropping by using the correlation of the GHZ triplet state.

\item Alice judges the Bob's result by the results of the particles
$P_1, P_3$ according to table I, and transfers her results to be consistent with Bob's results.

\item Alice and Bob obtain a sharing key by using the data sifting, the error correction and privacy amplification technologies.
\end{enumerate}
}

Let us now explain the above protocol in detail. In our protocol the sequence of the GHZ 
triplet states may be generated by one communicator, e.g. Alice or by a quantum source. 
For each 
GHZ triplet we let Alice have two particles and Bob have one particle from the GHZ 
triplet. The roles of the Alice's two particles are different: one is for the key 
and another is for setting up the correlation of the GHZ triplet. In steps 1 Alice only 
measures one of two GHZ particles, the result of this particle will be used 
for the quantum cryptographic key. The another particle will be measured in step 4. 
This particle is used to judge the Bob's result. In additional, by making use of
the particle $P_3$ Alice and Bob do not need to discard the sent 
quantum bits (qubits) that have been always discarded in all the previous protocols. 
This is the superiority of our protocol.

In step 4 Alice's measurement on particle $P_3$ must base on the 
measurement bases of the particles $P_1$ and $P_2$. The reason is that after Alice and Bob
measure respectively their particles $P_1$ and $P_2$, the state of the particle $P_3$
is determined by the correlation of the GHZ triplet state. If both Alice and Bob measure 
their particles ($P_1, P_2$) using the same measurement bases, i.e., $x$ or $y$ direction, 
Alice measures her particle $P_3$ along the $x$ direction, otherwise, Alice measures her
particle using the $y$ measurement basis. After measure Alice gains the result of the particle
$P_3$, which is any of the four states $\{|x+\rangle, |x-\rangle, |y+\rangle, |y-\rangle\}$.

Alice and Bob check the eavesdropping by using the correlation of the GHZ triplet state. 
If the results of the particles $P_1, P_2, P_3$ satisfy any of the Eqs.(7-10), which means
the results are in the table I with the correlation of the GHZ triplet state, we say the correlations among three particles are perfect. After transmission, Alice and Bob publicly compare the results of the GHZ triplet states on a sufficiently large random subset of the 
set of all GHZ triplets. If they find that the 
test subset is indeed perfectly correlated, they can refer that the remaining untested 
subset is also perfectly correlated, and therefore may be used for the quantum 
cryptographic key. The method is as follows: Bob send publicly Alice a random sub-sequence 
of his results, Alice compares the Bob's results with her corresponding results. If the
correlations of their results may be found in the table I, the correlation is perfect, 
otherwise it means the eavesdropping. For example, if the results of the particles $P_1, 
P_2, P_3$ are respectively the $|x+\rangle, |x-\rangle, |x-\rangle$, the results are 
perfect, if the results of the particles $P_1, P_2, P_3$ are respectively the 
$|x-\rangle, |x-\rangle, |x-\rangle$, Bob's particle must be eavesdropped by the 
eavesdropper or be disturbed by the noise.

It needs to stress that Alice's results must be consistent with Bob's results for getting 
the raw quantum key. In step 6, Alice can judge Bob's results by the results of the particles
$P_1$ and $P_3$, but Bob gains no knowledge on Alice's results, because Bob only has one 
result of the GHZ triplet state. In addition, from table I we see the most correlation results 
of particles $P_1, P_2, P_3$ are different. For example when Alice's results of two particles
$P_1, P_3$ are $|x-\rangle$ and $|x-\rangle$, according to table I the Bob's result should be 
$|x+\rangle$, obviously, the results of the particles $P_1$ and $P_2$ are different. 
For obtaining a same key, Alice's results need to be consistent with Bob's results. We may use
two methods to gain the same key. One is that Alice changes her qubits according to Bob's results. Another is that Alice transfers her qubits to binary bits according to Bob's results.
For example, assume Bob transfers his qubits using the aforehand appointment $|x+\rangle, |y+\rangle \rightarrow 0, |x-\rangle, |y-\rangle \rightarrow 1$, if Bob's result is 
$|y+\rangle$ and Alice's result is $|x-\rangle$, both Alice and Bob transfer their 
qubits to binary bits $0$. This way is different from the previous protocols, in which Alice 
and Bob have a same aforehand appointment.

As the previous protocols the raw quantum key distribution is useless in practice
because limited eavesdropping may be undetectable, yet it may leak some information, 
and errors are to be expected even in the absence of eavesdropping. For these reasons, 
our scheme needs to supplement some classical tools such as privacy amplification, error correction and data sifting, so we add the step 7 in our protocol. The implementation of 
these supplemented classic tools are the same as in the previous documents [9].

The important point is that the efficiency of the quantum key distribution has been 
improved in this protocol. Because 
Alice and Bob randomly measure their particles $P_1$ and $P_2$ before the Alice's 
measurement on the particle $P_3$, and Alice can measure the particle $P_3$ according 
to the measurement bases of the particles $P_1, P_2$, 
no Bob's qubits need to be discard in the proposed scheme.
This way largely improves the efficiency of 
the distribution for cryptographic key. Consider the quantum cryptographic system
will not be perfect because of the losing qubits ($l$) in the 
measurement and in the quantum channel, in order to be left with a key 
of $L$ qubits Alice should send $L'>(L+l)$ qubits to Bob. The efficiency is
\begin{equation}
\eta =\frac{L}{(L+l)}<100\%.
\end{equation}
In the previous protocol, the optimal efficiency is $\eta' ={L}/{2(L+l)}<50\%$. Obviously,
when the lossing qubits are same, $\eta>\eta'$. From Eq.(11), we see as $l\rightarrow 0$
$\eta\rightarrow 100\%$, it means that an ideal unsymmetrical GHZ protocol corresponds to
the $100\%$ efficiency. It need to stress that here the discard particles are the 
same as BB84 protocol for obtaining a $L$ qubits because the particle $P_3$ is finally 
discard in our scheme.

Let us now consider the security of this protocol. It is warranted by the correlation of 
the GHZ triplet state. From the table I it is clear that even if the eavesdropper knows what measurements Alice and Bob made (that is, $x$ or $y$) on particles $P_1, P_2, P_3$, she can 
only determine whether their results are the same or 
opposite and also that the eavesdropper will gain no knowledge of what Alice's and Bob's 
actually are, because the eavesdropper will has the probability of 1/2 of making a mistake. 
In fact, the eavesdropper can not gain any knowledge from Alice even if the measurement bases,
she can only obtain the Bob's measurement bases that is public. By this knowledge, the
eavesdropper can not obtain the useful information from Alice and Bob's sharing key.

The eavesdropper can not succeed to obtain the key by the intercept/resend attack defined in 
[9]. Suppose that the eavesdropper has managed to get a hold of Alice and Bob's key, 
she then intercepts Bob's particle $P_2$ and send another particle $P^*_2$ to Bob. In this
case, Alice's and the eavesdropper's three particles construct a GHZ triplet. Eavesdropper 
can not obtain the key, because the Alice's and Bob's particles are not a GHZ triplet, 
there are no correlated or anticorrelated result, eavesdropper's interception will introduce error and can be detected by Alice and Bob. 

The entanglement attacks is no use in our protocol. To show that, Let us assume that the eavesdropper has been able to entangle an ancilla in state $|A\rangle$ with the GHZ
triplet state that Alice and Bob are using. The state describing the state of the GHZ 
triplet and the ancilla is 
\begin{equation}
|\Psi\rangle=\frac{1}{\sqrt{2}}(|z+z+z+\rangle+|z-z-z-\rangle)\otimes|A\rangle.
\end{equation}
By using the $x$ and $y$ eigenstates and Eq.(7), The eavesdropper get

\begin{equation}
\begin{array}{rl}
U|\Psi\rangle=&\frac{1}{2}(|x+x+\rangle\otimes |x+\rangle\otimes |A_1\rangle
+|x-x-\rangle\otimes |x+\rangle\otimes |A_2\rangle\\
&\\
&+|x+x-\rangle\otimes |x-\rangle\otimes |A_3\rangle
+|x-x+\rangle\otimes |x-\rangle\otimes |A_4\rangle
).
\end{array}
\end{equation}
where $U$ denote the unitary transformation. By projecting the above state onto 
$|\phi\rangle=\alpha_1|x+x+\rangle+\alpha_2|x-x-\rangle+\alpha_3|x+x-\rangle
+\alpha_4|x-x+\rangle$, the eavesdropper creates the states
\begin{equation}
|\Psi\rangle_E=\frac{1}{2}
(|x+\rangle\otimes (\alpha^*_1|A_1\rangle+\alpha^*_2|A_2\rangle)
+|x-\rangle\otimes (\alpha^*_3|A_3\rangle+\alpha^*_4|A_4\rangle)
).
\end{equation}
If the eavesdropper can gain the Bob's qubits, she can obtain the key. However,
Eq.(14) shows that the eavesdropper can not obtain Bob's results. By the similar method, the eavesdropper can not obtain the Bob's qubits when Alice's and Bob's states satify Eqs.(8-10).

There is an additional points to notice here. Like all the previous schemes, our scheme 
itself can also not prevent the men-in-middle attack. 
The men-in-middle attack may be described as follows: When the legitimate communicator 
Alice communicates the legitimate communicator Bob, Eve 
intercepts all qubit sent by Alice, and communicates Bob with impersonating Alice. 
Finally, Eve obtains two keys $K_{AE}, K_{EB}$, where $K_{AE}$ represents the secret 
key between Alice and eavesdropper, and $K_{AE}$ represents the secret key between Bob and 
eavesdropper. As a result eavesdropper can easily decrypt the ciphertext sent by Alice or Bob. 
To prevent the men-in-middle attack, Alice and Bob may verify their identity by using the 
identity verification technologies [15]. This method need to share a aforehand key between Alice and Bob, the verification proceeding is similar to that presented in [15].

In conclusions, I propose a quantum key distribution protocol based on the GHZ triplet states, 
the security is warranted by the correlation of the GHZ triplet states. This protocol is unsymmetrical because Alice has two particles and Bob only has one particle from the GHZ 
triplet. The unsymmetrical 
characteristic let the proposed protocol have the following merits. 1) The efficiency of the quantum key distribution is excellent, it approaches $100\%$. This will be very useful in the practical application. 2) The security is higher than the previous protocol, because Alice's
bases for the quantum key are secret and Alice's and Bob's bases are unsymmetrical. 

I wish to thank Professor G. Guo and Dr. Z. Wang for their useful discussion on the GHZ states. This research was supported by the National Natural Science Foundation of China, Grants No. 69803008.

\vspace*{1.5cm}
\begin{flushleft}
References
\end{flushleft}

\begin{enumerate}
\item C. H. Bennett, and G.Brassard,  Advances in 
Cryptology: Proceedings of Crypto 84, August 1984, Springer - Verlag, p. 475. 
\item C. H. Bennett,  Phys. Rev. Lett., {\bf 68}, 3121, (1992). 
\item A. K.Ekert, Phys. Rev. Lett., {\bf 67}, 661, (1991). 
\item C.H. Bennett, G. Brassard, and N. D. Mermin, Phys. Rev. Lett. {\bf 68}, 557 (1992).
\item C.A.Fuchs, N.Gisin, R.B.Griffiths, C.S.Niu, and A.Peres, 
Phys. Rev. A, {\bf 56}, 1163, (1997).
\item B.A.Slutsky, R.Rao, P.C.Sun, and Y.Fainman, 
Phys. Rev. A, {\bf 57}, 2383, (1998).
\item H. E. Brandt, Phys. Rev. A {\bf 59}, 2665 (1999).
\item C. Niu, and R. Griffiths, Phys. Rev. A, {\bf 58}, 4377, (1998).
\item C. H. Bennett, F. Bessette, G. Brassard, L. Salvail and J. Smolin, J. Cryptology {\bf 5}, 
3 (1992). 
\item J.Breguet, A.Muller, and N.Gisin, Journal of Modern Optics, {\bf 41}, 2405, (1994).
\item C. Marand and P.D.Townsend, Optics Lett. {\bf 20},  
1695 (1995).
\item W.T.Buttler, R.J.Hughes, P.G.Kwiat, et. al., Phys. Rev. A {\bf 57}, 2379 (1998).
\item A.Einstein, B.Podolsky, and N.Rosen, Phys. Rev. {\bf 47}, 777 (1935).
\item D. Greenberger, M. A. Horne, and A. Zeilinger, in Bell's Theorem, Quantum theory, and Conceptions of Universe, edited by M.Kaftos (Kluwer Academic, Dordrecht, 1989); 
D. Greenberger, M. A. Horne, A. Shimony, and A. Zeilinger, Am. J. Phys. 
{\bf 58}, 1131 (1990).
\item G. Zeng and W. Zhang, Phys. Rev. A, {\bf 61}, 2000 (In press).
\end{enumerate}

\end{document}